\documentclass[11pt,a4paper]{article}
\usepackage{amsfonts,amssymb}
\usepackage{cite}
\usepackage[T2A]{fontenc}
\usepackage[cp1251]{inputenc}
\usepackage[english]{babel}
\usepackage{amsmath}%
\usepackage[unicode]{hyperref}

\newtheorem{theorem}{Theorem}[section]
\newtheorem{definition}{Definition}[section]

\newtheorem{remark}{Remark}[section]
\newtheorem{conjecture}{Conjecture}[section]

\numberwithin{equation}{section}

\title{\bf Generalized invariant manifolds \\ for integrable equations and their applications.}
\author{\bf I.T. Habibullin, A.R. Khakimova, A.O. Smirnov.}
\date{}

\begin{document}
\maketitle

\begin{abstract}
In the article we discuss the notion of the generalized invariant manifold introduced in our previous study. In the literature the method of the differential constraints is well known as a tool for constructing particular solutions for the nonlinear partial differential equations. Its essence is in adding to the nonlinear PDE, a much simpler, as a rule ordinary, differential equation, compatible with the given one. Then any solution of the ODE is a particular solution of the PDE as well. However the main problem is to find this compatible ODE. Our generalization is that we look for an ordinary differential equation that is compatible not with the nonlinear partial differential equation itself, but with its linearization.  Such a generalized invariant manifold is effectively sought. Moreover, it allows one to construct such important attributes of integrability theory as Lax pairs and recursion operators for integrable nonlinear equations. In this paper, we show that they provide a way to construct particular solutions to the equation as well.
\end{abstract}

\maketitle


\section{Introduction}

In the article a notion of the generalized invariant manifold for nonlinear integrable equation is discussed. Recently in our works \cite{HKP2016JPA}--\cite{HK2020JPA} it has been observed that this kind objects provide an effective tool for evaluating the Lax pairs  and recursion operators.

The approach developed in \cite{HKP2016JPA}--\cite{HK2020JPA} explains the essence of the Lax pair phenomenon. In fact, the Lax pair in $1+1$ dimension is naturally (internally) derived from the nonlinear equation under consideration. First we find the linearization (Fr$\acute{e}$chet derivative) of the nonlinear equation. The linearized equation obviously includes the dynamical variables of the original equation as well, which are here considered as functional parameters. Now we find an ordinary differential equation compatible with the linearized equation, which also depends on the dynamical variables of the original equation. We call this ordinary differential equation a generalized invariant manifold. For a given equation, there are many such manifolds, including nonlinear ones. In order to evaluate the generalized invariant manifold we use the consistency with the linearized equation which allows to derive a system of differential (difference) equations that is highly overdetermined due to the presence of the independent parameters -- dynamical variables of the original nonlinear equation. In all of the examples discussed in \cite{HKP2016JPA}--\cite{HK2020JPA} (KdV, Kaup-Kupershmidt equation, Krichever-Novikov equation, Volterra type lattices from Yamilov list, two equations of KdV type found by Svinolupov and Sokolov, Garifullin-Mikhailov-Yamilov non-autonomous lattice, sine-Gordon equation and several hyperbolic type equations, etc.) the corresponding overdetermined systems are effectively solved and the desired non-trivial manifolds are found. Trivial generalized invariant manifolds are constructed quite elementarily by using the classical or higher symmetries (see examples in \cite{Khakimova18}). A manifold that is consistent  with the linearized equation if and only if the original nonlinear equation is satisfied is called non-trivial. Actually, this requirement means that a pair consisting of the linearized equation and the generalized invariant manifold defines a Lax pair. It is curious that usual Lax pairs do not belong to this class, but they can be derived from properly chosen nonlinear generalized invariant manifolds by suitable transformations. Note that new Lax pairs are interesting in themselves. For instance, a generalized invariant manifold generated by a consistent pair of linear invariant manifolds is easily transformed into the recursion operator. In [6] it was shown with the example of the Volterra lattice that a nonlinear Lax pair can be used for constructing particular solutions of the nonlinear equation. 

Let's briefly discuss the content of the article. In the second section we recall the definition of the invariant manifold and generalized invariant manifold for the differential equations in partial derivatives. We explane how to look for the generalized invariant manifold and convince the reader why it can  be effectively found. We hypothesize that any integrable equation admits a consistent pair of linear invariant manifolds and give examples confirming the hypothesis. We assert, based on our previous work, that consistent pairs of linear invariant manifolds can be used to construct both recursion operators and Lax pairs.  We illustrate the algorithm with the examples of NLS system and  mKdV equation in \S 3-\S 5. The consistent pair of the linear generalized invariant manifolds usually can be reduced to nonlinear one of smaller order. In this form, the invariant manifold provides an efficient way to derive the Dubrovin equations, from which finite-gap solutions are obtained (about method of finite-gap integration see \cite{DubrovinMatveevNovikov}--\cite{Smirnov1995}).  The description of the spectral curve, the derivation and study of the Dubrovin equations for the NLS equation are presented in \S 3.1-\S 3.3. The corresponding solutions of the generalized invariant manifolds and their relation with the Novikov equation are considered in \S 3.4. Examples of one-phase and two-phase solutions of the NLS equation are given in \S 3.5, \S 3.6. Derivation of the Dubrovin equations for mKdV equation is represented in \S 4.

\section{Invariant manifolds and their generalization}

The concept of an invariant manifold is well known in the theory of partial differential equations. It forms the basis of the method of differential constraints, widely used to construct particular solutions of nonlinear equations. We recall briefly the main points of the method of the invariant manifolds using the example of equations of evolutionary type
\begin{equation}
u_t=f(x,t,u,u_x,u_{xx},\dots,u_k), \quad u_j=\frac{\partial^ju}{\partial x^j}.\label{main}
\end{equation}
An ordinary differential equation of the order $r$
\begin{equation}
u_r=g(x,t,u,u_x,u_{xx},\dots,u_{r-1})\label{im}
\end{equation}
is called an invariant manifold for the equation \eqref{main} if it is consistent  with \eqref{main},  or, in other words, if the following condition is met
\begin{equation}
\left.D^r_xf-D_tg\right|_{\eqref{main},\eqref{im}}=0.\label{mainim}
\end{equation}
Here $D_x$ and $D_t$ are operators of the total derivative with respect to $x$ and correspondingly, to $t$. 

It is clear that if a solution $u(x,t)$ of the equation \eqref{main} for some moment $t=t_0$ satisfies the equation \eqref{im}, then it remains a solution of \eqref{im} at all values of time t. This is the invariance of the equation \eqref{im}.

Obviously  relation \eqref{mainim} defines a PDE for the desired  function $g$. Sometimes this equation can be solved explicitly, although in the general case the problem of finding the function $g$ is rather complicated.

The situation radically changes if we look for an ordinary differential equation that is compatible not with the nonlinear equation \eqref{main} itself, but with its linearization
\begin{equation}
U_t=\frac{\partial f}{\partial u}U+\frac{\partial f}{\partial u_x}U_x+\frac{\partial f}{\partial u_{xx}}U_{xx}+\dots+\frac{\partial f}{\partial u_k}U_k.\label{linmain}
\end{equation}

Let's move on to a precise definition. Consider the ordinary differential equation of the form
\begin{equation}
U_m=F(x,t,U,U_x,U_{xx},\dots,U_{m-1};u,u_x,u_{xx},\dots,u_n),\label{oim}
\end{equation}
where $U=U(x,t)$ is a sought function, while an arbitrary solution $u=u(x,t)$ of the original equation \eqref{main}, is interpreted in \eqref{oim} as a functional parameter. Actually, the variables  $x$, $t$, $U$, $U_x$, $U_{xx}$, \dots, $U_{m-1}$, $u$, $u_x$, $u_{xx}$, \dots, $u_n$ in \eqref{oim} are independent variables.

\begin{definition}\label{def1} Equation \eqref{oim} determines a generalized invariant manifold if the relation
\begin{equation}
\left.D^m_xU_t-D_tU_m\right|_{\eqref{main},\eqref{linmain},\eqref{oim}}=0\label{cond_linoim}
\end{equation}
is satisfied identically for all values of the variables $\left\{u_j\right\}$, $x$, $t$, $U$, $U_x$, \dots, $U_{m-1}$.
\end{definition}

Here the variables $u_t$, $U_t$ as well as their derivatives with respect to $x$ are expressed due to the equations \eqref{main} and \eqref{linmain}, the variables $U_m, U_{m+1}, \dots$ are replaced by means of \eqref{oim}. To emphasize that the solution $u(x,t)$ is arbitrary, we consider the variables $u$, $u_x$, $u_ {xx}, \dots$ as independent ones. By virtue of this assumption, the problem of finding the function  $F(x,t,U,U_x,U_{xx},\dots,U_{m-1};u,u_x,u_{xx},\dots,u_n)$ is overdetermined and as  it is approved by numerous examples is effectively solved. 

Linear GIM i.e. generalized invariant manifolds of the form 
\begin{align*}
LU=0, 
\end{align*}
where $L$ is a linear differential operator 
\begin{equation*}
L=\sum^{N}_{i=0}a_i(u,u_x,u_{xx},\ldots)D_x^i
\end{equation*}
are of the special interest.
\begin{definition}
Let equations $L_1U=0$ and $L_2U=0$ define linear generalized invariant manifolds for the equation \eqref{main}. We refer to these two manifolds consistent if for any $\lambda, \mu \in C$ the linear combination 
\begin{align*}
\left(\lambda L_1+\mu L_2\right)U=0
\end{align*}
is a generalized invariant manifold for \eqref{main}.
\end{definition}

The following hypothesis is supported by numerous examples (see \cite{HKP2016JPA}--\cite{HK2020JPA}).

\begin{conjecture}  \label{conjecture}
Equation \eqref{main} is integrable if and only if it admits a pair of the consistent linear generalized invariant manifolds such that the ratio 
\begin{align*}
R=L_1^{-1}L_2
\end{align*}
is a pseudodifferential operator (which is in fact the recursion operator for \eqref{main}).
\end{conjecture}
Examples can be found below in \S 3.2 and at the end of \S 5.

\section{Invariant manifolds for the NLS equation}

Now in this section we evaluate an invariant manifold of the first order (the simplest nontrivial!) for the nonlinear Schr\"odinger equation, that follows from the system
\begin{equation}
\begin{aligned}\label{NLS}
&iu_t=u_{xx}+2u^2v,\\
&iv_t=-v_{xx}-2v^2u
\end{aligned}
\end{equation}
under appropriate additional condition. Let us first determine the linearized equation for the system due to the rule \eqref{linmain}.
\begin{equation}
\begin{aligned}\label{linNLS}
&iU_t=U_{xx}+4uvU+2u^2V,\\
&iV_t=-V_{xx}-2v^2U-4uvV.
\end{aligned}
\end{equation}
According to the Definition~\ref{def1} generalized invariant manifold is a system of the ordinary differential equations compatible with \eqref{linNLS} for arbitrary solution $u=u(x,t),$ $v=v(x,t)$ of \eqref{NLS}. We look for it in the form
\begin{equation}
\begin{aligned}\label{oimNLS}
&U_x=f(U,V,u,v),\\
&V_x=g(U,V,u,v).
\end{aligned}
\end{equation}
The consistency condition for the equations \eqref{linNLS} and \eqref{oimNLS} gives an overdetermined system of equations for a pair of unknowns $f$ and $g$ that is effectively solved and defines a generalized invariant manifold given by a system of the form (for the details see Appendix below)
\begin{equation}
\begin{aligned}\label{xsysN}
&U_x=\lambda U -2u\sqrt{C-UV},\\
&V_x=-\lambda V -2v\sqrt{C-UV},
\end{aligned}
\end{equation}
where $\lambda$ and $C$ are arbitrary constants. Due to the obtained equations the linearized equation \eqref{linNLS} converts into a system of the ordinary differential equations:
\begin{equation}\label{tsysN}
\begin{aligned}
&iU_t=(2uv+\lambda^2)U-2(u_x+\lambda u)\sqrt{C-UV},\\
&iV_t=-(2uv+\lambda^2)V+2(v_x-\lambda v)\sqrt{C-UV}.
\end{aligned}
\end{equation}
The following statement is easily approved by a direct computation.

\begin{theorem} A pair of systems \eqref{xsysN} and \eqref{tsysN} is compatible if and only if the functions $u$ and $v$ solve equation \eqref{NLS}.
\end{theorem}

Therefore the pair of equations \eqref{xsysN} and \eqref{tsysN} defines a Lax pair for the NLS equation. Unlike the usual Lax pair found by V.E. Zakharov and A.B. Shabat, this pair is nonlinear and contains two arbitrary constants, but with the help of a simple technique it is reduced to the usual one \cite{ZakharovShabat}. Indeed by setting $C=0$, $U=\varphi^2$, $V=\psi^2$ we reduce equations \eqref{xsysN} and \eqref{tsysN} to the form 
\begin{equation*}
\begin{aligned}
&\varphi_x=\frac{1}{2}\lambda \varphi -iu\psi,\\
&\psi_x=-iv \varphi -\frac{1}{2}\lambda\psi,
\end{aligned}
\end{equation*}
and, respectively,
\begin{equation*}
\begin{aligned}
&\varphi_t=\left(uv+\frac{1}{2}\lambda^2\right) \varphi -i(u_x+\lambda u)\psi,\\
&\psi_t=i(v_x-\lambda v) \varphi -\left(uv+\frac{1}{2}\lambda^2\right)\psi.
\end{aligned}
\end{equation*}

\subsection{Invariant manifolds and spectral curves}

Let us show that the found nonlinear Lax pair is interesting in itself, as it provides  opportunities for building particular  solutions of the NLS equation.
Let us change the variables in the nonlinear Lax pair as follows $U=u\Phi$, $V=v\Psi$ and bring it to the form
\begin{equation}
\begin{aligned}\label{xsysNN}
&\frac{u_x}{u}\Phi+\Phi_x -\lambda\Phi=-2\sqrt{C-\Phi\Psi uv},\\
&\frac{v_x}{v}\Psi+\Psi_x +\lambda\Psi=-2\sqrt{C-\Phi\Psi uv}
\end{aligned}
\end{equation}
and 
\begin{equation}
\begin{aligned}
&i\frac{u_t}{u}\Phi+i\Phi_t=(2uv+\lambda^2)\Phi-2\left(\frac{u_x}{u}+\lambda\right)\sqrt{C-\Phi\Psi uv}, \label{tsysNN}\\
&i\frac{v_t}{v}\Psi+i\Psi_t=-(2uv+\lambda^2)\Psi+2\left(\frac{v_x}{v}-\lambda\right)\sqrt{C-\Phi\Psi uv}.
\end{aligned}
\end{equation}
Assume that parameters $C$ and $\lambda$ are related to each other in such a way that $C$ is a polynomial of $\lambda$ with constant coefficients:
\begin{equation}
C=\frac{1}{4}\prod_{k=1}^{2N+2}(\lambda-\lambda_k)=\frac{1}{4}\nu^2(\lambda)\label{C}
\end{equation}
and look for solutions to the nonlinear Lax equations in the form
\begin{equation}
\Phi=\prod_{k=1}^{N}(\lambda-\gamma_k),\quad \Psi=-\prod_{k=1}^{N}(\lambda-\beta_k).\label{phipsi}
\end{equation}
We note that equality \eqref{C} defines the equation for the spectral hyperelliptic curve of the $N$-gap solution of the NLS equation (see \cite{KotlyarovIts}-\cite{Smirnov1995}).

Now we substitute representations \eqref{C} and \eqref{phipsi} into system \eqref{xsysNN} and compare the coefficients before the power $\lambda^{N}$ and derive relations (well known trace formulae)
\begin{equation}
\begin{aligned}\label{relations}
&\frac{u_x}{u}=-\sum_{k=1}^N \gamma_k+\frac{1}{2}\sum_{k=1}^{2N+2} \lambda_k,\\
&\frac{v_x}{v}=\sum_{k=1}^N \beta_k-\frac{1}{2}\sum_{k=1}^{2N+2} \lambda_k.
\end{aligned}
\end{equation}
After substituting the polynomials \eqref{phipsi} into system \eqref{xsysNN}
and taking $\lambda=\gamma_j$ in the first equation and $\lambda=\beta_j$ in the second we  get the well known Dubrovin's formulae
\cite{DubrovinMatveevNovikov}
\begin{equation}\label{xdubr}
\gamma_j'=\frac{\nu(\gamma_j)}{\prod_{k\neq j}(\gamma_j-\gamma_k)}, \quad
\beta_j'=-\frac{\nu(\beta_j)}{\prod_{k\neq j}(\beta_j-\beta_k)},
\end{equation}
where $\gamma'_j=\frac{d\gamma_j}{dx}$, $\beta'_j=\frac{d\beta_j}{dx}$. By applying the same manipulations to \eqref{tsysNN} we obtain
\begin{equation}
\begin{aligned}\label{tdubr}
&i\dot{\gamma_j}=\frac{\left(-\sum_{k\neq j} \gamma_k+\frac{1}{2}\sum_{k=1}^{2N+2} \lambda_k\right)\nu(\gamma_j)}{\prod_{k\neq j}(\gamma_j-\gamma_k)}, \\
&i\dot{\beta_j}=-\frac{\left(-\sum_{k\neq j} \beta_k+\frac{1}{2}\sum_{k=1}^{2N+2} \lambda_k\right)\nu(\beta_j)}{\prod_{k\neq j}(\beta_j-\beta_k)},
\end{aligned}
\end{equation}
where $\dot{\gamma_j}=\frac{d\gamma_j}{dt}$, $\dot{\beta_j}=\frac{d\beta_j}{dt}$.
In order to get the focusing NLS equation
\begin{equation*}
iu_t=u_{xx}+2\left|u\right|^2u
\end{equation*}
we assign to the system \eqref{NLS} a constraint of the form $v=\bar{u}$ where the bar over a letter means the complex conjugation. Then solution $\left(\Phi,\Psi\right)$ to the nonlinear Lax pair \eqref{xsysNN}, \eqref{tsysNN} can be chosen in such a way 
\begin{equation*}
\bar{\Phi}\left(-\bar{\lambda}\right)=(-1)^{N+1}\Psi(\lambda).
\end{equation*}
Function $C(\lambda)$ and parameters $\lambda_j$, $\beta_j$, $\gamma_j$ satisfy the involution 
\begin{equation*}
C(\lambda)=\bar{C}\left(-\bar{\lambda}\right), \quad \bar{\lambda}_j=-\lambda_j, \quad \beta_j=-\bar{\gamma}_j.
\end{equation*}
Evolution of $\gamma_j$ in $x$ and $t$ is determined by a pair of the systems of ordinary differential equations
\begin{align}
&\gamma_j'=\frac{\nu(\gamma_j)}{\prod_{k\neq j}(\gamma_j-\gamma_k)}, \label{xsystem}\\
&i\dot{\gamma_j}=\frac{\left(-\sum_{k\neq j} \gamma_k+\frac{1}{2}\sum_{k=1}^{2N+2} \lambda_k\right)\nu(\gamma_j)}{\prod_{k\neq j}(\gamma_j-\gamma_k)}. \label{tsystem}
\end{align}
Thus we arrive at systems of ODE describing the well-known algebro-geometric solutions for the NLS equations (see \cite{KotlyarovIts}).

\selectlanguage{english}

\begin{theorem} A pair of systems \eqref{xsystem} and \eqref{tsystem} is compatible.
\end{theorem}

\noindent
\textbf{Proof.}
Let us show that a pair of systems \eqref{xsystem}, \eqref{tsystem} is compatible. To do this, we differentiate the system \eqref{xsystem} with respect to $t$, multiply by $i$ and subtract from it the system \eqref{tsystem} differentiated with respect to $x$. We get

\begin{equation}\label{sys_comp}
\begin{aligned}
i\frac{d}{d t}\left(\gamma_j'\right)-\frac{d}{d x}\left(i\dot{\gamma_j}\right)= &\frac{i\frac{d}{d t}\nu(\gamma_j)}{\prod_{k\neq j}(\gamma_j-\gamma_k)}-\frac{i\nu(\gamma_j)\frac{d}{d t}\prod_{k\neq j}(\gamma_j-\gamma_k)}{\prod_{k\neq j}(\gamma_j-\gamma_k)^2}\\
&+\sum_{s\neq j}\left(\gamma_s'\right)\frac{\nu(\gamma_j)}{\prod_{k\neq j}(\gamma_j-\gamma_k)}\\
&-\left(-\sum_{k\neq j} \gamma_k+\frac{1}{2}\sum_{k=1}^{2N+2} \lambda_k\right)\frac{\frac{d}{d x}\nu(\gamma_j)}{\prod_{k\neq j}(\gamma_j-\gamma_k)}\\
&+\left(-\sum_{k\neq j} \gamma_k+\frac{1}{2}\sum_{k=1}^{2N+2} \lambda_k\right)\frac{\nu(\gamma_j)\frac{d}{d x}\prod_{k\neq j}(\gamma_j-\gamma_k)}{\prod_{k\neq j}(\gamma_j-\gamma_k)^2}.
\end{aligned}
\end{equation}

Let us find all of the derivatives
\begin{align*}
i\frac{d}{d t}\nu(\gamma_j), \qquad \frac{d}{d x}\nu(\gamma_j), \qquad i\frac{d}{d t}\prod_{k\neq j}(\gamma_j-\gamma_k), \qquad \frac{d}{d x}\prod_{k\neq j}(\gamma_j-\gamma_k) 
\end{align*}
separately:

\begin{equation*}
\begin{aligned}
i\frac{d}{d t}\nu(\gamma_j)=&\frac{i\dot{\gamma_j}\nu(\gamma_j)}{2}\sum_{k=1}^{2N+2}\frac{1}{\gamma_j-\lambda_k}\\
=&\frac{1}{2}\left(-\sum_{k\neq j} \gamma_k+\frac{1}{2}\sum_{k=1}^{2N+2} \lambda_k\right)\frac{\nu^2(\gamma_j)}{\prod_{k\neq j}(\gamma_j-\gamma_k)}\sum_{k=1}^{2N+2}\frac{1}{\gamma_j-\lambda_k},
\end{aligned}
\end{equation*}
similarly
\begin{equation*}
\begin{aligned}
\frac{d}{d x}\nu(\gamma_j)=&\frac{\gamma_j'\nu(\gamma_j)}{2}\sum_{k=1}^{2N+2}\frac{1}{\gamma_j-\lambda_k}
=\frac{1}{2}\frac{\nu^2(\gamma_j)}{\prod_{k\neq j}(\gamma_j-\gamma_k)}\sum_{k=1}^{2N+2}\frac{1}{\gamma_j-\lambda_k},
\end{aligned}
\end{equation*}
and then
\begin{equation*}
\begin{aligned}
i\frac{d}{d t}\prod_{k\neq j}(\gamma_j-\gamma_k)=&\prod_{k\neq j}(\gamma_j-\gamma_k)\sum_{s\neq j}\frac{i\dot{\gamma_j}-i\dot{\gamma_s}}{\gamma_j-\gamma_s}\\
=&\left(-\sum_{k\neq j} \gamma_k+\frac{1}{2}\sum_{k=1}^{2N+2} \lambda_k\right)\nu(\gamma_j)\sum_{s\neq j}\frac{1}{\gamma_j-\gamma_s}\\
&+\prod_{k'\neq j}(\gamma_j-\gamma_{k'})\sum_{s\neq j}\left(\sum_{k\neq s}\gamma_k\right)\frac{\nu(\gamma_s)}{(\gamma_j-\gamma_s)\prod_{k\neq s}(\gamma_s-\gamma_k)}\\
&-\frac{1}{2}\sum_{k=1}^{2N+2} \lambda_k \prod_{k'\neq j}(\gamma_j-\gamma_{k'}) \sum_{s\neq j}\frac{\nu(\gamma_s)}{(\gamma_j-\gamma_s)\prod_{k\neq s}(\gamma_s-\gamma_k)},
\end{aligned}
\end{equation*}
finally we get a representation for the last of them
\begin{equation*}
\begin{aligned}
\frac{d}{d x}\prod_{k\neq j}(\gamma_j-\gamma_k)=&\prod_{k\neq j}(\gamma_j-\gamma_k)\sum_{s\neq j}\frac{\gamma_j'-\gamma_s'}{\gamma_j-\gamma_s}\\
=&\nu(\gamma_j)\sum_{s\neq j}\frac{1}{\gamma_j-\gamma_s}-\prod_{k\neq j}(\gamma_j-\gamma_k) \sum_{s\neq j}\frac{\nu(\gamma_s)}{(\gamma_j-\gamma_s)\prod_{k\neq s}(\gamma_s-\gamma_k)}.
\end{aligned}
\end{equation*}
We substitute the obtained equalities into \eqref{sys_comp} and after simple transformations we arrive at
\begin{equation}\label{sys_comp2}
\begin{aligned}
i\frac{d}{d t}\left(\gamma_j'\right)-\frac{d}{d x}\left(i\dot{\gamma_j}\right)=&\frac{\nu(\gamma_j)}{\prod_{k\neq j}(\gamma_j-\gamma_k)}\left[-\sum_{s\neq j}\left(\sum_{k\neq s}\gamma_k\right)\frac{\nu(\gamma_s)}{(\gamma_j-\gamma_s)\prod_{k\neq s}(\gamma_s-\gamma_k)}\right.\\
&\left.+\sum_{s\neq j}\frac{\nu(\gamma_s)}{\prod_{k\neq s}(\gamma_s-\gamma_k)}+\sum_{k\neq j}\gamma_k\sum_{s\neq j}\frac{\nu(\gamma_s)}{(\gamma_j-\gamma_s)\prod_{k\neq s}(\gamma_s-\gamma_k)}\right].
\end{aligned}
\end{equation}
We notice that the first two members in the bracket are simplified as
\begin{align*}
-\sum_{s\neq j}\left(\sum_{k\neq s}\gamma_k\right)&\frac{\nu(\gamma_s)}{(\gamma_j-\gamma_s)\prod_{k\neq s}(\gamma_s-\gamma_k)}+\sum_{s\neq j}\frac{\nu(\gamma_s)}{\prod_{k\neq s}(\gamma_s-\gamma_k)}\\
=&\sum_{s\neq j}\left(\gamma_j-\gamma_s-\sum_{k\neq s}\gamma_k\right)\frac{\nu(\gamma_s)}{(\gamma_j-\gamma_s)\prod_{k\neq s}(\gamma_s-\gamma_k)}\\
=&\sum_{s\neq j}\left(-\sum_{k\neq j}\gamma_k\right)\frac{\nu(\gamma_s)}{(\gamma_j-\gamma_s)\prod_{k\neq s}(\gamma_s-\gamma_k)}.
\end{align*}
Then, by taking into account the obtained relation, we conclude that equality \eqref{sys_comp2} takes the form
\begin{align*}
i\frac{d}{d t}\left(\gamma_j'\right)-\frac{d}{d x}\left(i\dot{\gamma_j}\right)=0.
\end{align*}
The theorem is proved.

\subsection{Consistent pair of linear invariant manifolds}

Here we present an example approving the Conjecture~\ref{conjecture}. From the nonlinear invariant manifold (\ref{xsysN}) we derive a consistent pair of linear invariant manifolds for the system (\ref{NLS}). To this end we differentiate both equations in (\ref{xsysN}) with respect to $x$
\begin{align*}
&U_{xx}=\lambda U_x -2u_x\sqrt{C-UV}+u\frac{U_xV+V_xU}{\sqrt{C-UV}},\\
&V_{xx}=-\lambda V_x -2v_x\sqrt{C-UV}+v\frac{U_xV+V_xU}{\sqrt{C-UV}}
\end{align*}
and then exclude irrationalities in the obtained equations due to relations:
\begin{align*}
&\frac{U_xV+V_xU}{\sqrt{C-UV}}=-2(uV+vU),\\ 
&\sqrt{C-UV}=\frac{\lambda U-U_x}{2u}=\frac{-\lambda V-V_x}{2v}.
\end{align*}
As a result we obtain a linear relation 
\begin{equation}\label{Linear}
L_2W=\lambda L_1W,
\end{equation}
where $W=(U,V)^T$ and the operators are as follows
\begin{equation*}\label{L1}
L_1=\left(\begin{array}{cc}
D_x-\frac{u_x}{u}&0 \\
0&-D_x+\frac{v_x}{v} 
\end{array} \right),
\end{equation*}
and 
\begin{equation*}\label{L2}
L_2=\left(\begin{array}{cc}
D_x^2-\frac{u_x}{u}D_x+2uv&2u^2 \\
2v^2&D_x^2-\frac{v_x}{v}D_x+2uv 
\end{array} \right).
\end{equation*}
It is approved straightforwardly that the linear constraint defined by equation \eqref{Linear} is compatible with the linearized equation \eqref{linNLS}, i.e. it defines a linear generalized invariant manifold with a parameter $\lambda$ for the NLS equation. It is easily verified that a pseudo differential operator  $L_1^{-1}L_2$ coincides with the recursion operator for the NLS system \eqref{NLS}
\begin{equation*}\label{RO}
R=\left(\begin{array}{cc}
D_x+2uD_x^{-1}v&2uD_x^{-1}u \\
-2vD_x^{-1}v&-D_x-2vD_x^{-1}u 
\end{array} \right).
\end{equation*}

\subsection{Integrals of the systems}

The overdetermined system of equations \eqref{xsystem}, \eqref{tsystem} admits integrals of the form
\begin{align}
&\sum_{j=1}^{N}\int_{\gamma_{j}(0,0)}^{\gamma_{j}(x,t)}\gamma^k\frac{d\gamma}{\nu(\gamma)}=0,\quad k=0,1,...,N-3, \label{integralsNLS1}\\
&\sum_{j=1}^{N}\int_{\gamma_{j}(0,0)}^{\gamma_{j}(x,t)}\gamma^{N-2}\frac{d\gamma}{\nu(\gamma)}=-it, \label{integralsNLS2}\\
&\sum_{j=1}^{N}\int_{\gamma_{j}(0,0)}^{\gamma_{j}(x,t)}\gamma^{N-1}\frac{d\gamma}{\nu(\gamma)}=x-\frac{i}{2}t\sum_{k=1}^{2N+2}\lambda_k \label{integralsNLS3}
\end{align}
that are derived directly from the systems by using some elementary manipulations and subsequent integration.

\subsection{Novikov equation}

Let us first express coefficients of the polynomial $P=\left(\Phi,\Psi\right)^T$ in terms of the solution $(u,v)$ of the NLS equation \eqref{NLS} obtained due to \eqref{relations}. To this end we rewrite equation \eqref{Linear} in a convenient form
\begin{align}\label{RP}
\widetilde{R}P=\lambda P,
\end{align}
where
\begin{align*}
\widetilde{R}=\left(\begin{array}{cc} \frac{1}{u} & 0 \\ 0 & \frac{1}{v} \end{array}\right)R\left(\begin{array}{cc} u & 0 \\ 0 & v \end{array}\right).
\end{align*}

We introduce notations for the coefficients of the polynomial $P$
\begin{align}\label{P}
P=\left(\begin{array}{c} 1 \\ -1 \end{array}\right)\lambda^N + \left(\begin{array}{c} r_1 \\ s_1 \end{array}\right)\lambda^{N-1}+\ldots+\left(\begin{array}{c} r_N \\ s_N \end{array}\right).
\end{align}
By virtue of the expansion \eqref{P} equation \eqref{RP} gives rise to 
\begin{equation*}
\begin{aligned}
\widetilde{R}\left(\left(\begin{array}{c} 1 \\ -1 \end{array}\right)\lambda^N + \left(\begin{array}{c} r_1 \\ s_1 \end{array}\right)\lambda^{N-1}+\ldots+\left(\begin{array}{c} r_N \\ s_N \end{array}\right)\right)\\
=\left(\begin{array}{c} 1 \\ -1 \end{array}\right)\lambda^{N+1} + \left(\begin{array}{c} r_1 \\ s_1 \end{array}\right)\lambda^{N}+\ldots+\left(\begin{array}{c} r_N \\ s_N \end{array}\right)\lambda.
\end{aligned}
\end{equation*}
Comparison of the coefficients at $\lambda^{N+1}$ in \eqref{NLS} yields 
\begin{align}\label{RNp1}
\widetilde{R}\left(\begin{array}{c} 0 \\ 0 \end{array}\right)=\left(\begin{array}{c} 1 \\ -1 \end{array}\right).
\end{align}
We recall that the operator $\widetilde{R}$ contains integration. It is easily checked that for appropriate choice of the constants of integration equation \eqref{RNp1} is satisfied.

Then we pass to the coefficients at $\lambda^{N}$ and get 
\begin{align*}
\widetilde{R}\left(\begin{array}{c} 1 \\ -1 \end{array}\right)=\left(\begin{array}{c} r_1 \\ s_1\end{array}\right)
\end{align*}
that implies
\begin{align*}
{r_1 \choose s_1}= {\frac{1}{u} \left(u_x+c_1u\right) \choose \frac{1}{v}\left(v_x-c_1v\right)}.
\end{align*}
By continuing this process we find for $k\geq 1$:
\begin{align*}
{r_k \choose s_k}= {\frac{1}{u}\left(g_k+c_1g_{k-1}+\ldots+c_k u\right) \choose \frac{1}{v}\left(h_k+c_1h_{k-1}+\ldots+c_k (-v)\right)},
\end{align*}
where $c_i$,  $i=\overline{1,k}$ are arbitrary constants, the vector $\left(g_j,h_j\right)$ coincides with the generator of the homogeneous symmetry of the order $k$ 
\begin{align*}
u_{\tau_j}=g_j, \quad v_{\tau_j}=h_j
\end{align*}
of the NLS system \eqref{NLS}. Finally, comparing the coefficients in front of $\lambda^0$ we find
\begin{align*}
r_{N+1}=0, \quad s_{N+1}=0
\end{align*}
that actually coincides with the Novikov equation
\begin{align*}
&g_N+c_1g_{N-1}+\ldots+c_N u=0, \\
&h_N+c_1h_{N-1}+\ldots+c_N (-v)=0.
\end{align*}

\subsection{Examples.}

Below we present two examples illustrating the use of the Dubrovin equations \eqref{xsystem}, \eqref{tsystem} by taking $N=1$ and $N=2$.

\medskip

\noindent
\textbf{Example 1.} In the particular case when $N=1$ and 
\begin{align*}
\nu(\gamma)=(\gamma-\lambda_1)(\gamma-\lambda_2) \quad \mbox{with} \quad \lambda_1=\eta+i\xi, \quad \lambda_2=-\eta+i\xi
\end{align*}
we get a system of compatible equations for determining the unknown $\gamma=\gamma(x,t)$:
\begin{align*}
&\gamma'=(\gamma-\lambda_1)(\gamma-\lambda_2), \\
&i\dot{\gamma}=2i\xi(\gamma-\lambda_1)(\gamma-\lambda_2),
\end{align*}
which is easily solved and gives
\begin{align*}
\gamma=\eta \tanh(2\xi \eta t+\eta x+s_0)+i\xi.
\end{align*}
In order to find $u=u(x,t)$ we first solve the equation 
\begin{align*}
\frac{u_x}{u}=-\gamma+2i\xi.
\end{align*}
Integration of this equation yields 
\begin{align*}
u=\frac{e^{i\xi x A(t)}}{\cosh(2\xi \eta t+\eta x+s_0)}.
\end{align*}
We substitute the obtained ansatz into the NLS equation $iu_t=u_{xx}+2\left|u\right|^2u$ and find 
\begin{align*}
A(t)=\eta e^{i(\eta^2-\xi^2)t+i\varphi_0}
\end{align*}
and finally get the well-known soliton solution 
\begin{align*}
u(x,t)=\frac{\eta e^{i\left(\xi x+(\eta^2-\xi^2)t+\varphi_0\right)}}{\cosh(2\xi \eta t+\eta x+s_0)}.
\end{align*}

\medskip

\noindent
\textbf{Example 2.} Let us take $N=2$ and assume that the hyperelliptic curve is as follows
\begin{align*}
\nu(\lambda)=(\lambda^2-4)^3.
\end{align*}
In order to look for functions $\gamma_1(x,t)$ and $\gamma_2(x,t)$ we use the integrals \eqref{integralsNLS1}-\eqref{integralsNLS3}, which in this case take the form:
\begin{align*}
&\sum_{j=1}^{2}\int_{\gamma_{j}(0,0)}^{\gamma_{j}(x,t)}\frac{d\gamma}{\nu(\gamma)}=-it,\\
&\sum_{j=1}^{2}\int_{\gamma_{j}(0,0)}^{\gamma_{j}(x,t)}\gamma\frac{d\gamma}{\nu(\gamma)}=x.
\end{align*}
These integrals are evaluated in a closed form and generate a system of algebraic equations for $\gamma_1(x,t)$, $\gamma_2(x,t)$, which is easily solved and gives rise to
\begin{align*}
\gamma_j(x,t)=-2 \frac{X(1-iT)+(-1)^j(iT-X-1)(iT+X-1)R}{(X^2+T^2+1)(X^2+T^2+4iT-3)}, \quad j=1,2,
\end{align*}
where $T=4t$,  $X=2x$ and
\begin{align*}
R=\sqrt{(T^2+2iT+X^2+3)(X-iT+1)(X+iT-1)}.
\end{align*}
Then we find $u(x,t)$ according to \eqref{relations}:
\begin{align*}
u(x,t)=\left(1-\frac{4(1-iT)}{X^2+T^2+1}\right)e^{-2it}.
\end{align*}
Obviously the obtained solution coincides with the well known two-phase Peregrine soliton \cite{Peregrine}.

\section{Invariant manifolds for the mKdV equation (complex valued case)}

A complex valued version of the modified KdV equations has important physical applications (see \cite {Demon11})
\begin{equation}
\begin{aligned}\label{mKdV_complexv}
&u_\tau+u_{xxx}+6|u|^2u_x=0.
\end{aligned}
\end{equation}
The equation is obtained by imposing an involution of the form $v=\bar u$, where the bar over the letter means complex conjugation, in the system of equations
\begin{equation}
\begin{aligned}\label{mKdV_complex}
&u_\tau+u_{xxx}+6uvu_x=0,\\
&v_\tau+v_{xxx}+6uvv_x=0.
\end{aligned}
\end{equation}
Linearization of \eqref{mKdV_complex} leads to a system
\begin{equation}
\begin{aligned}\label{mKdV_lin}
&U_\tau+U_{xxx}+6uvU_x+6vu_xU+6uu_xV=0,\\
&V_\tau+V_{xxx}+6uvV_x+6vu_xU+6uu_xV=0.
\end{aligned}
\end{equation}
It is noteworthy that the generalized invariant manifold \eqref{xsysN} found in the previous section for the NLS equation, is also a generalized invariant manifold for the mKdV equation \eqref{mKdV_complex}.
It is not surprising since the systems \eqref{NLS} and \eqref{mKdV_complex} commute with each other. It can be easily approved that system 
\begin{equation}
\begin{aligned}\label{OIM_mKdV}
&U_x=\lambda U -2u\sqrt{C-UV},\\
&V_x=-\lambda V -2v\sqrt{C-UV}
\end{aligned}
\end{equation}
is consistent with \eqref{mKdV_lin} for any solution $(u,v)$ of \eqref{mKdV_complex}. Due to \eqref{OIM_mKdV} the linearized equation \eqref{mKdV_lin} is reduced to the form
\begin{equation}
\begin{aligned}\label{mKdVlin}
&U_\tau=2\left(u_{xx}+2u^2v+\lambda u_x+\lambda^2 u\right)\sqrt{-UV+C}-\left(2uv_x-2vu_x-2\lambda uv-\lambda^3\right)U,\\
&V_\tau=2\left(v_{xx}+2uv^2-\lambda v_x+\lambda^2 v\right)\sqrt{-UV+C}-\left(2uv_x-2vu_x-2\lambda uv-\lambda^3\right)V.
\end{aligned}
\end{equation}
It is easily approved that \eqref{OIM_mKdV}, \eqref{mKdVlin} define a nonlinear Lax pair for the system \eqref{mKdV_complex}.

Since the systems \eqref{OIM_mKdV} and \eqref{mKdVlin} are very similar to those studied in the previous section (see \eqref{xsysN}, \eqref{tsysN}) we investigate them by the same way. At first we change the variables by taking 
$U=u\Phi$, $V=v\Psi$ and get
\begin{equation}
\begin{aligned}\label{xsysmKdV}
&\frac{u_x}{u}\Phi+\Phi_x -\lambda\Phi=-2\sqrt{C-\Phi\Psi uv},\\
&\frac{v_x}{v}\Psi+\Psi_x +\lambda\Psi=-2\sqrt{C-\Phi\Psi uv}
\end{aligned}
\end{equation}
and 
\begin{equation}\label{tsysmKdV}
\begin{aligned}
&\frac{u_\tau}{u}\Phi+\Phi_\tau=2\left(\frac{u_{xx}}{u}+2uv+\lambda \frac{u_x}{u}+\lambda^2\right)\sqrt{C-\Phi\Psi uv}-\left(2uv_x-2vu_x-2\lambda uv-\lambda^3\right)\Phi, \\
&\frac{v_\tau}{v}\Psi+\Psi_\tau=2\left(\frac{v_{xx}}{v}+2uv-\lambda \frac{v_x}{v}+\lambda^2\right)\sqrt{C-\Phi\Psi uv}-\left(2uv_x-2vu_x-2\lambda uv-\lambda^3\right)\Psi.
\end{aligned}
\end{equation}
We use the same spectral curve:
\begin{equation}
C=\frac{1}{4}\prod_{k=1}^{2N+2}(\lambda-\lambda_k)=\frac{1}{4}\nu^2(\lambda)\label{Cm}
\end{equation}
and look for solutions to the nonlinear Lax equations in the same form
\begin{equation}
\Phi=\prod_{k=1}^{N}(\lambda-\gamma_k),\quad \Psi=-\prod_{k=1}^{N}(\lambda-\beta_k).\label{phipsim}
\end{equation}

Afterwards we substitute representations \eqref{Cm} and \eqref{phipsim} into system \eqref{xsysmKdV} and by comparing the coefficients before the power $\lambda^{N}$ we derive the trace formulae \eqref{relations}. The next step is to substitute the polynomials \eqref{phipsim} into system \eqref{xsysmKdV}. Then we set $\lambda=\gamma_j$ in the first equation and $\lambda=\beta_j$ in the second and get the system of ordinary differential equations defining the dynamics of the roots on $x$  
\cite{DubrovinMatveevNovikov}
\begin{equation}\label{xdubrm}
\gamma_j'=\frac{\nu(\gamma_j)}{\prod_{k\neq j}(\gamma_j-\gamma_k)}, \quad
\beta_j'=-\frac{\nu(\beta_j)}{\prod_{k\neq j}(\beta_j-\beta_k)},
\end{equation}
where $\gamma'_j=\frac{d\gamma_j}{dx}$, $\beta'_j=\frac{d\beta_j}{dx}$. Let us derive equations describing the time evolution of the functions $\gamma(x,t)$, $\beta(x,t)$. To this end we substitute explicit representations of the functions $\Phi$, $\Psi$ and $C$ into \eqref{tsysmKdV} and set $\lambda=\gamma_j$ in the first equation and $\lambda=\beta_j$ in the second:
\begin{equation}
\begin{aligned}\label{tdubrm0}
&\dot{\gamma_j}=\left(\frac{u_{xx}}{u}+2uv+\lambda \frac{u_x}{u}+\lambda^2\right)\frac{\nu(\gamma_j)}{\prod_{k\neq j}(\gamma_j-\gamma_k)}, \\
&\dot{\beta_j}=-\left(\frac{v_{xx}}{v}+2uv-\lambda \frac{v_x}{v}+\lambda^2\right)\frac{\nu(\beta_j)}{\prod_{k\neq j}(\beta_j-\beta_k)},
\end{aligned}
\end{equation}
where we used notations $\dot{\gamma_j}=\frac{d\gamma_j}{d\tau}$, $\dot{\beta_j}=\frac{d\beta_j}{d\tau}$.
To get a closed system for $\gamma_j$, $\beta_j$ we exclude $\frac{u_x}{u}$, $\frac{v_x}{v}$, $\frac{v_{xx}}{v}$ and $\frac{u_{xx}}{u}$ due to \eqref{relations}. For the term $uv$ we deduce two equations
\begin{align*}
&4uv=2\sum_{k\neq j} \gamma_k'+\sum_{k=1}^{2N+2} \lambda_k\sum_{k\neq j} \gamma_k-2\sum_{k\neq s} \gamma_k\gamma_s-2\sum_{k=1}^{N} \left(\gamma_k\right)^2+\frac{1}{4}\left(\sum_{k=1}^{2N+2} \lambda_k\right)^2-\sum_{k\neq s} \lambda_k \lambda_s,\\
&4uv=-2\sum_{k\neq j} \beta_k'+\sum_{k=1}^{2N+2} \lambda_k\sum_{k\neq j} \beta_k-2\sum_{k\neq s} \beta_k\beta_s-2\sum_{k=1}^{N} \left(\beta_k\right)^2+\frac{1}{4}\left(\sum_{k=1}^{2N+2} \lambda_k\right)^2-\sum_{k\neq s} \lambda_k \lambda_s
\end{align*}
by comparing coefficients at the power $\lambda^{N+1}$ in \eqref{tsysmKdV}. As a result system \eqref{tdubrm0} converts into a system of equations describing the time evolution of the roots
\begin{equation}
\begin{aligned}\label{tdubrm}
&\dot{\gamma_j}=\left(-\frac{1}{2}\sum_{k=1}^{2N+2} \lambda_k\sum_{k\neq j} \gamma_k+\sum_{k\neq s\neq j} \gamma_k\gamma_s+\frac{3}{8}\left(\sum_{k=1}^{2N+2} \lambda_k\right)^2-\frac{1}{2}\sum_{k\neq s} \lambda_k \lambda_s\right)\frac{\nu(\gamma_j)}{\prod_{k\neq j}(\gamma_j-\gamma_k)}, \\
&\dot{\beta_j}=-\left(\frac{1}{2}\sum_{k=1}^{2N+2} \lambda_k\sum_{k\neq j} \beta_k-\sum_{k\neq s\neq j} \beta_k\beta_s-\frac{3}{8}\left(\sum_{k=1}^{2N+2} \lambda_k\right)^2+\frac{1}{2}\sum_{k\neq s} \lambda_k \lambda_s\right)\frac{\nu(\beta_j)}{\prod_{k\neq j}(\beta_j-\beta_k)}.
\end{aligned}
\end{equation}

It can be proved by a direct computation that systems \eqref{xdubrm} and \eqref{tdubrm} commute with one another.

In what follows, we impose on the  system \eqref{mKdV_complexv} reductions of  two types. 
Complex reduction $v=\bar{u}$ in \eqref{mKdV_complexv} is related to the involution 
\begin{align*}
\bar{\Phi}\left(-\bar{\lambda}\right)=(-1)^{N+1}\Psi(\lambda)
\end{align*}
of the eigenfunctions, that generates conditions for the zeros $\bar{\gamma}_k=-\beta_j$ and $\bar{\lambda}_k=-\lambda_j$ of the polynomials $\Phi$, $\Psi$, $C$, such that 
\begin{align*}
C(\lambda)=\prod_{j=1}^{N+1}\left(\lambda^2-2\lambda\, Re \lambda_j +\left|\lambda_j\right|^2\right).
\end{align*}
In the case of the reduction $v=u$ we obtain 
\begin{align*}
\Phi\left(-\lambda\right)=(-1)^{N+1}\Psi(\lambda)
\end{align*}
and $\gamma_k=-\beta_j$, $\lambda_k=-\lambda_j$, such that 
\begin{align*}
C(\lambda)=\prod_{j=1}^{N+1}\left(\lambda^2-\lambda_k^2\right).
\end{align*}
Obviously both reductions are compatible with the dynamics \eqref{xdubrm}, \eqref{tdubrm}.
By using the found solution $\{\gamma_j\}_{j=1}^N$ of the Dubrovin equations one can find solution $u$ of the equation \eqref{mKdV_complexv} due to the formula \eqref{relations}. Such kind solutions have earlier been studied in \cite{Smirnov2016}.

The overdetermined system of equations \eqref{xdubrm}, \eqref{tdubrm} admits integrals of the form
\begin{align}
&\sum_{j=1}^{N}\int_{\gamma_{j}(0,0)}^{\gamma_{j}(x,\tau)}\gamma^k\frac{d\gamma}{\nu(\gamma)}=0,\quad k=0,1,...,N-4, \label{integralsmKdV1}\\
&\sum_{j=1}^{N}\int_{\gamma_{j}(0,0)}^{\gamma_{j}(x,\tau)}\gamma^{N-3}\frac{d\gamma}{\nu(\gamma)}=\tau,\label{integralsmKdV2}\\
&\sum_{j=1}^{N}\int_{\gamma_{j}(0,0)}^{\gamma_{j}(x,\tau)}\gamma^{N-2}\frac{d\gamma}{\nu(\gamma)}=\left(\frac{1}{2}\sum_{k=1}^{2N+2}\lambda_k\right)\tau, \label{integralsmKdV3}\\
&\sum_{j=1}^{N}\int_{\gamma_{j}(0,0)}^{\gamma_{j}(x,\tau)}\gamma^{N-1}\frac{d\gamma}{\nu(\gamma)}=x+\left(\frac{3}{8}\left(\sum_{k=1}^{2N+2} \lambda_k\right)^2-\frac{1}{2}\sum_{k\neq s} \lambda_k \lambda_s\right)\tau. \label{integralsmKdV4}
\end{align}

Taking into account the dependence on $t$ and $\tau$ in the equations \eqref{integralsNLS1}-\eqref{integralsNLS3} and \eqref{integralsmKdV1}-\eqref{integralsmKdV4}, we have
\begin{align*}
&\sum_{j=1}^{N}\int_{\gamma_{j}(0,0,0)}^{\gamma_{j}(x,t,\tau)}\gamma^k\frac{d\gamma}{\nu(\gamma)}=0,\quad k=0,1,...,N-4,\\
&\sum_{j=1}^{N}\int_{\gamma_{j}(0,0,0)}^{\gamma_{j}(x,t,\tau)}\gamma^{N-3}\frac{d\gamma}{\nu(\gamma)}=\tau,\\
&\sum_{j=1}^{N}\int_{\gamma_{j}(0,0,0)}^{\gamma_{j}(x,t,\tau)}\gamma^{N-2}\frac{d\gamma}{\nu(\gamma)}=-it+\left(\frac{1}{2}\sum_{k=1}^{2N+2}\lambda_k\right)\tau,\\
&\sum_{j=1}^{N}\int_{\gamma_{j}(0,0,0)}^{\gamma_{j}(x,t,\tau)}\gamma^{N-1}\frac{d\gamma}{\nu(\gamma)}=x-\left(\frac{i}{2}\sum_{k=1}^{2N+2}\lambda_k\right)t+\left(\frac{3}{8}\left(\sum_{k=1}^{2N+2} \lambda_k\right)^2-\frac{1}{2}\sum_{k\neq s} \lambda_k \lambda_s\right)\tau.
\end{align*}

Note that  formulas \eqref{integralsNLS1}-\eqref{integralsNLS3} and \eqref{integralsmKdV1}-\eqref{integralsmKdV4} can be easily generalized to solutions which simultaneously satisfy several equations from the AKNS hierarchy. In this case we obtain the following representation for the integrals
\begin{align*}
\sum_{j=1}^{N}\int_{\gamma_{j}(0,0,...)}^{\gamma_{j}(t_0,t_1,t_2,...)}\gamma^{N-k}\frac{d\gamma}{\nu(\gamma)}=t_k +
\sum_{j>k} c_{k,j}t_j, \quad k\geq 1,
\end{align*}
where $t_1=x$, $t_2=-it$, $t_3=\tau$ and $t_j$, for $j>3$ correspond to higher symmetries. Here $c_{k,j}$ are constant coefficients.

\section{Invariant manifolds for the mKdV equation (real case)}

In this section we look for the generalized invariant manifold for the mKdV equation which can be obtained from (\ref{mKdV_complex}) by imposing the real involution $u=v$:
\begin{align}\label{mKdV}
u_t=u_{xxx}+6u^2u_x
\end{align}
and then show how to derive from it the Lax pair, recursion operator and a consistent pair of the linear invariant manifolds. 

First, we linearize equation \eqref{mKdV}:
\begin{align}\label{linmKdV}
U_t=U_{xxx}+6u^2U_x+12uu_xU.
\end{align}
Since equation \eqref{mKdV} is reduced to the equation with cubic nonlinearity
\begin{align*}
w_t=w_{xxx}+2w^3
\end{align*}
by substitution $u=w_x$ the linearizations of these two equations are related by a similar replacement. Indeed if we put $U=W_x$, then arrive at the equation 
\begin{align}\label{linmKdV_W}
W_t=W_{xxx}+6u^2W_x
\end{align}
that is much simpler than \eqref{linmKdV}. Hence it is more convenient to work with this one. Below we will search an ODE of the form 
\begin{align*}
W_{xx}=F(W_x,W,u),
\end{align*}
compatible with linear equation \eqref{linmKdV_W} for any solution $u(x,t)$ of the equation \eqref{mKdV}. Omitting the computations we represent only the answer
\begin{align}\label{nonlinmKdVx}
W_{xx}=2u\sqrt{-W^2_x+\lambda W^2+C}+\lambda W,
\end{align}
where $\lambda$ and $C$ are arbitrary constants. By virtue of the found equation, linearization \eqref{linmKdV_W} turns into the form 
\begin{align}\label{nonlinmKdVt}
W_{t}=\left(\lambda+2u^2\right)W_x+2u_x\sqrt{-W^2_x+\lambda W^2+C}.
\end{align}
It worth mentioning that the found nonlinear equations provide a Lax pair for \eqref{mKdV}, or more precisely
\begin{theorem} A pair of equations \eqref{nonlinmKdVx} and \eqref{nonlinmKdVt} are compatible if and only if the function $u$ solves equation \eqref{mKdV}.
\end{theorem}

\begin{remark} We note that generalized invariant manifolds \eqref{OIM_mKdV} and \eqref{nonlinmKdVx} are related by the following change of the variables. We put in \eqref{nonlinmKdVx} $\lambda=\xi^2$ and introduce $U$, $V$ in such a way
\begin{align*}
U=W_x-\xi W, \quad V=W_x+\xi W.
\end{align*}
Then we get
\begin{align*}
&U_x=\xi U -2u\sqrt{C-UV},\\
&V_x=-\xi V -2v\sqrt{C-UV}.
\end{align*}
\end{remark}

Let us reduce pair of nonlinear equations \eqref{nonlinmKdVx}, \eqref{nonlinmKdVt} for the case when $C=0$ to the usual Lax pair of equation \eqref{mKdV}. To this end we change the variables in the following way
\begin{align*}
W=2\varphi\psi, \quad W_x=\sqrt{\lambda}\left(\varphi^2+\psi^2\right)
\end{align*}
then in the new variables the equation \eqref{nonlinmKdVx} converts into a system of linear equations
\begin{align*}
&\varphi_x=-iu\varphi+\frac{1}{2}\sqrt{\lambda}\psi,\\
&\psi_x=\frac{1}{2}\sqrt{\lambda}\varphi+iu\psi
\end{align*}
and similarly \eqref{nonlinmKdVt} turns into
\begin{align*}
&\varphi_t=-i\left(u_{xx}+2u^3+u\lambda\right)\varphi+\frac{1}{2}\sqrt{\lambda}\left(2iu_x+2u^2+\lambda\right)\psi,\\
&\psi_t=-\frac{1}{2}\sqrt{\lambda}\left(2iu_x-2u^2-\lambda\right)\varphi+i\left(u_{xx}+2u^3+u\lambda\right)\psi.
\end{align*}

In order to bring it to a standard form we make a replacement
\begin{equation*}\label{Laxstand}
\Phi=\left(\begin{array}{cc}
1&1 \\
1&-1 
\end{array} \right)\tilde\Phi,
\end{equation*}
where $\Phi=(\varphi,\psi)^T$ and $\tilde\Phi=(\tilde\varphi,\tilde\psi)^T$. Then we get 
\begin{align*}
&\tilde\varphi_x=\xi\tilde\varphi-iu\tilde\psi,\\
&\tilde\psi_x=-iu\tilde\varphi-\xi\tilde\psi
\end{align*}
and 
\begin{align*}
&\tilde\varphi_t=\left(4\xi^3+2u^2\xi\right)\tilde\varphi-i\left(4u\xi^2+2u_x\xi+u_{xx}+2u^2\right)\tilde\psi,\\
&\tilde\psi_t=-i\left(4u\xi^2-2u_x\xi+u_{xx}+2u^2\right)\tilde\varphi-\left(4\xi^3+2u^2\xi\right)\tilde\psi,
\end{align*}
where $\lambda=4\xi^2$.

In the equality \eqref{nonlinmKdVx}, we get rid of irrationality by squaring and then rewrite the result  as
\begin{align*}
\frac{W^2_{xx}}{u^2}-2\lambda\frac{WW_{xx}}{u^2}+\lambda^2\frac{W^2}{u^2}+4W_x^2-4\lambda W^2-4C=0.
\end{align*}
Then we differentiate the obtained equation with respect to $x$. The found equation turns out to be linear
\begin{align*}
W_{xxx}-\frac{u_x}{u}W_{xx}+4u^2W_x=\lambda\left(W_x-\frac{u_x}{u}W\right).
\end{align*}
Actually it defines a linear invariant manifold, compatible with the equation \eqref{linmKdV_W}. Let us rewrite it in the form
\begin{align}\label{R2}
L_2W=\lambda L_1W,
\end{align}
where
\begin{align*}
L_1=D_x-\frac{u_x}{u}, \quad L_2=D^3-\frac{u_x}{u}D_{xx}+4u^2D_x.
\end{align*}
Since $U=W_x$ from \eqref{R2} we get a relation 
\begin{align*}
L_2D_x^{-1}U=\lambda L_1D_x^{-1}U
\end{align*}
for the solution $U$ of the linearization \eqref{linmKdV} of the mKdV equation \eqref{mKdV}. The latter allows to get the recursion operator for \eqref{mKdV}:
\begin{align}
R=D_xL_1^{-1}L_2D_x^{-1}=D_x^2+4u^2+4u_xD_x^{-1}u.
\end{align}

\section{Appendix}

Let us show how equations \eqref{xsysN} were constructed. The consistency condition for equations \eqref{linNLS} and \eqref{oimNLS}, mentioned above, is written in the form:
\begin{equation}\label{condNLS}
\begin{aligned}
&\left.\frac{\partial}{\partial x}\left(U_{xx}+4uvU+2u^2V\right)-i\frac{\partial}{\partial t}\left(f(U,V,u,v)\right)\right|_{\eqref{NLS},\eqref{linNLS},\eqref{oimNLS}}=0,\\
&\left.\frac{\partial}{\partial x}\left(-V_{xx}-2v^2U-4uvV\right)-i\frac{\partial}{\partial t}\left(g(U,V,u,v)\right)\right|_{\eqref{NLS},\eqref{linNLS},\eqref{oimNLS}}=0.
\end{aligned}
\end{equation}
We will consider variables $u$, $v$ and their derivatives with respect to $x$ and $U$, $V$ to be independent. The variables $u_t$, $v_t$, $U_t$, $V_t$, $D^k_xU$, $D^k_xV$, where $k\geq 1$, in equalities \eqref{condNLS}, will be excluded by virtue of equations \eqref{NLS},\eqref{linNLS} and \eqref{oimNLS}. Then we get an overdetermined system of equations of the form:
\begin{align*}
&F(U,V,u,v,u_x,v_x,v_{xx})=0, \\
&G(U,V,u,v,u_x,v_x,u_{xx})=0.
\end{align*}
In the obtained relations, we will equate the coefficients of independent variables $u_x$, $v_x$, $u_{xx}$, $v_{xx}$ and step by step determine the form of the sought functions $f(U,V,u,v)$ and $g(U,V,u,v)$. 

Let us compare the coefficients at the highest derivatives of functions $u$ and $v$, i.e. at $u_{xx}$ and $v_{xx}$, as well as at $u^2_x$ and $v^2_x$, then we get:
\begin{align*}
&f(U,V,u,v)=f_1(U,V)u+f_2(U,V), \\
&g(U,V,u,v)=g_1(U,V)v+g_2(U,V).
\end{align*}
Note that the dependence of functions $f(U,V,u,v)$ and $g(U,V,u,v)$ on variables $u$ and $v$ is defined, so we can equate the coefficients at these variables. 

Analyzing the equations obtained by comparing the coefficients for $u_x$, $v_x$, $u$ and $v$, we determine the form of functions $f_1(U,V)$, $f_2(U,V)$ and $g_1(U,V)$, $g_2(U,V)$:
\begin{align*}
&f_1(U,V)=\sqrt{-4UV+c_1}, \quad f_2(U,V)=c_2U+c_3, \\
&g_1(U,V)=\sqrt{-4UV+c_4}, \quad g_2(U,V)=c_5V+c_6.
\end{align*}
By considering the remaining equations, we obtain a relationship between the constant parameters $c_i$, $i=\overline{1,6}$:
\begin{align*}
c_4=c_1, \quad c_5=-c_2, \quad c_3=c_6=0.
\end{align*}
Thus, functions $f(U,V,u,v)$ and $g(U,V,u,v)$ have the form:
\begin{align*}
&f(U,V,u,v)=\lambda U -2u\sqrt{C-UV}, \\
&g(U,V,u,v)=-\lambda V -2v\sqrt{C-UV},
\end{align*}
where $\lambda=c_2$, $C=\frac{1}{4}c_1$.

\section*{Conclusions.}

In the article we discussed the notion of the generalized invariant manifold for the nonlinear partial differential equations. We evaluated such manifolds for the nonlinear Scr\"odinger equation and the modified Korteweg-de Vries equation. We illustrated that this object provides an effective tool for constructing the recursion operator and the Lax pair. We have shown that the well-known Dubrovin equation, which is an important element of the  finite-zone integration method, can be easily derived from a suitably selected generalized invariant manifold.

\end{document}